\documentclass[apj]{emulateapj}
\usepackage{mathptmx}

\def\gtorder{\mathrel{\raise.3ex\hbox{$>$}\mkern-14mu
             \lower0.6ex\hbox{$\sim$}}}
\def\ltorder{\mathrel{\raise.3ex\hbox{$<$}\mkern-14mu
             \lower0.6ex\hbox{$\sim$}}}

\slugcomment{M85 transient 2006-1}

\shorttitle{M85\,OT\,2006-1}

\shortauthors{Ofek et al.}

\begin{document}

\title{The Environment of M85 optical transient 2006-1: constraints on the progenitor age and mass}
\author{
E.~O.~Ofek\altaffilmark{1},
S.~R.~Kulkarni\altaffilmark{1},
A.~Rau\altaffilmark{1},
S.~B.~Cenko\altaffilmark{1},
E.~W.~Peng\altaffilmark{2},
J.~P.~Blakeslee\altaffilmark{3},
P.~C\^ot\'e\altaffilmark{2},
L.~Ferrarese\altaffilmark{2},
A.~Jord\'an\altaffilmark{4},
S.~Mei\altaffilmark{5},
T.~Puzia\altaffilmark{2},
L.~D.~Bradley\altaffilmark{6},
D.~Magee\altaffilmark{7},
R.~Bouwens\altaffilmark{7},
}
\altaffiltext{1}{Division of Physics, Mathematics and Astronomy, California Institute of Technology, Pasadena, CA 91125, USA}
\altaffiltext{2}{Herzberg Institute of Astrophysics, Dominion Astrophysical Observatory, Victoria, BC, Canada V9E 2E7}
\altaffiltext{3}{Department of Physics and Astronomy, Washington State University, Pullman, WA}
\altaffiltext{4}{European Southern Observatory, Karl-Schwarzschild-Strasse 2, 85748, Garching, Germany}
\altaffiltext{5}{GEPI, Observatoire de Paris, Section de Meudon, Meudon Cedex, France}
\altaffiltext{6}{Department of Physics and Astronomy, The Johns Hopkins University, Baltimore, MD 21218}
\altaffiltext{7}{Department of Astronomy and Astrophysics, University of California, Santa Cruz, CA 95064}

\begin{abstract}

M85 optical transient 2006-1 (M85\,OT\,2006-1)
is the most luminous member of the small family of
V838~Mon-like objects,
whose nature is still a mystery.
This event took place
in the Virgo cluster of galaxies and
peaked at an absolute magnitude of $M_{I}\approx-13$.
Here we present Hubble Space Telescope images of
M85\,OT\,2006-1 and its environment,
taken before and after the eruption,
along with a spectrum of the host galaxy
at the transient location.
We find that the progenitor of M85\,OT\,2006-1
was not associated with any star forming region.
The $g$ and $z$-band absolute magnitudes of the progenitor
were fainter than about $-4$ and $-6$~mag, respectively.
Therefore, we can set a lower limit of $\sim50$\,Myr
on the age of the youngest stars at the location
of the progenitor that corresponds to a mass of $<7$~M$_{\odot}$.
Previously published line indices suggest
that M85 has a mean stellar age of $1.6\pm0.3$~Gyr.
If this mean age is representative of the progenitor
of M85\,OT\,2006-1, then
we can further constrain its mass to be less than $2$~M$_{\odot}$.
We compare the energetics and mass limit derived for
the M85\,OT\,2006-1 progenitor
with those expected from a simple model
of violent stellar mergers.
Combined with further modeling, these new clues may ultimately
reveal the true nature of these puzzling events.

\end{abstract}


\section{Introduction}
\label{Introduction}

M85 Optical Transient 2006-1 (M85\,OT\,2006-1;
J122523.82$+$181056.2) was discovered on 2006 Jan 6
by the Lick observatory supernova search
team (Filippenko et al. 2001\footnote{http://astro.berkeley.edu/$\sim$bait/kait.html})
as a faint, $V\sim19.3$~mag transient in the galaxy M85
(NGC\,4382), which is at a distance of $17.8$~Mpc (Mei et al. 2007).
Subsequent spectroscopy, as well as visible light and infra-red (IR)
photometry, presented in Kulkarni et al. (2007),
showed that M85\,OT\,2006-1
has a recession velocity of $880\pm130$~km~s$^{-1}$,
and is therefore associated with M85.
Moreover, we showed that
the temporal and spectral properties of this object
are unlike those of supernovae, novae, or luminous blue variables.

M85\,OT\,2006-1 peaked at absolute $I$-band magnitude of about $-13$.
The light curve settled into a $\sim60$~day plateau,
followed by a decrease in bolometric luminosity during which the
black-body emission peak shifted toward near-IR wavelengths.
The early spectrum of M85\,OT\,2006-1, obtained six weeks after discovery,
resembles that of a $\sim4600$~K black body,
with H$\alpha$ and H$\beta$ narrow emission lines
(full width at half maximum of $\sim350\,$km\,s$^{-1}$),
along with several other unidentified emission
lines.
%\footnote{Electronic version of the spectrum is available
%from http://astro.caltech.edu/$\sim$eran/Transients/M85OT/}.
Spitzer IR observations obtained about six months after the discovery
revealed a $\sim1000\,$K black body spectral energy distribution
(Rau et al. 2007).

The spectral and temporal properties of this object resemble
those of M31-RV
(discovered by Rich et al. 1989;
e.g., Mould et al. 1990; Bryan \& Royer 1992),
V838~Mon (discovered by Brown 2002; e.g.,
Kimeswenger et al. 2002; Bond et al. 2003; Corradi \& Munari 2007),
and possibly the less studied object V4332~Sgr (Martini et al. 1999).
However, the M85 transient is
the most luminous member of the V838~Mon class.
The favored model for this emerging class of V838~Mon-like objects
(also known as luminous red novae\footnote{this term was introduced by Kulkarni et al. 2007.})
is that they are the result of stellar mergers (e.g., Soker \& Tylenda 2006).
However, other models have been suggested to explain
these objects (e.g., Retter \& Marom 2003; Lawlor 2005).
The nature  of these events, with their energetics lying between
the realms of supernovae and novae, remains uncertain.

In this paper, we present Hubble Space Telescope (HST)-
Advanced Camera for Surveys (ACS)/Wide Field Camera (WFC) and 
Near Infrared Camera and Multi-Object Spectrometer
(NICMOS) observations,
as well as Palomar 5\,m spectroscopy,
of the environment of M85\,OT\,2006-1.
The observations are used
to characterize the environment of the transient
and to set a limit on the mass of the progenitor.

\section{Observations}
\label{Obs}

\begin{figure*}
\centerline{\includegraphics[width=17cm]{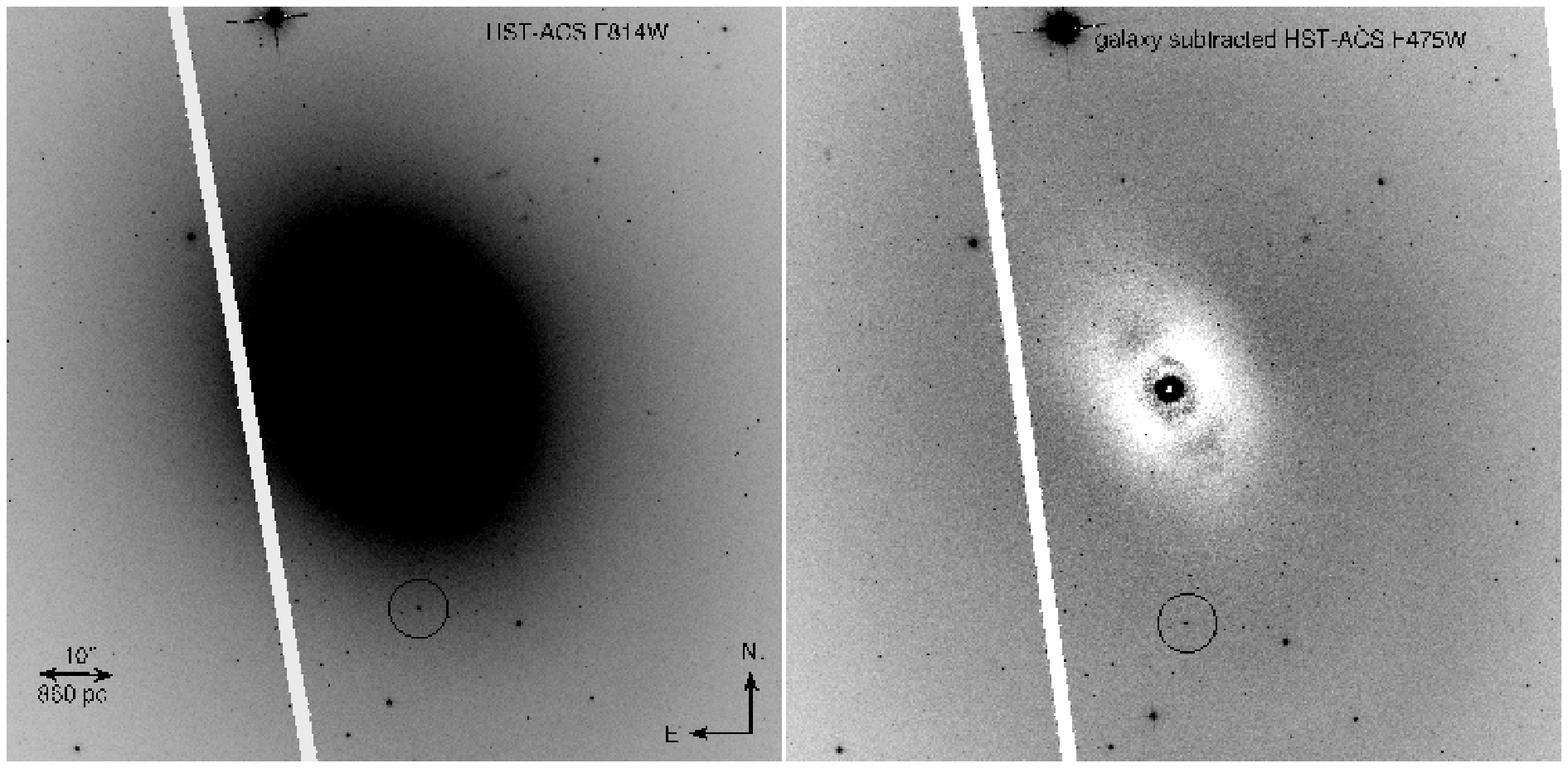}}
\caption{{\bf Left:} HST/ACS $F814W$-band image of the galaxy M85,
obtained on 2006 Jan 24 ($18$~days after the discovery).
The transient, which is well detected, is marked by a circle.
{\bf Right:} HST/ACS $F475W$-band image of M85 after subtraction
of the best fit Sersic model (using GalFit; Peng et al. 2002).
The subtracted model parameters are: 
effective radius $389''$;
Sersic index $3.0$;
axis ratio ($b/a$) $0.765$;
position angle $29.3^{\circ}$;
diskiness $-0.064$.
A different set of structural parameters is obtained when analyzing
the azimuthally averaged profile (Ferrarese et al. 2006).
We note, that the rough galaxy subtracted image allows us to show the
lack of dust and structure in the neighborhood of the transient.
Note that the gray scale level stretch in the left panel is about $5.2$ times
larger than in the right panel. The white band in the images is due to the gap
between the ACS CCDs.
The slit of the Palomar 5\,m telescope spectrum (Fig.~\ref{M85_2d_spec})
passes through the transient location and the center of the galaxy.
\label{f1}}
\end{figure*}

\begin{figure*}
\centerline{\includegraphics[width=17cm]{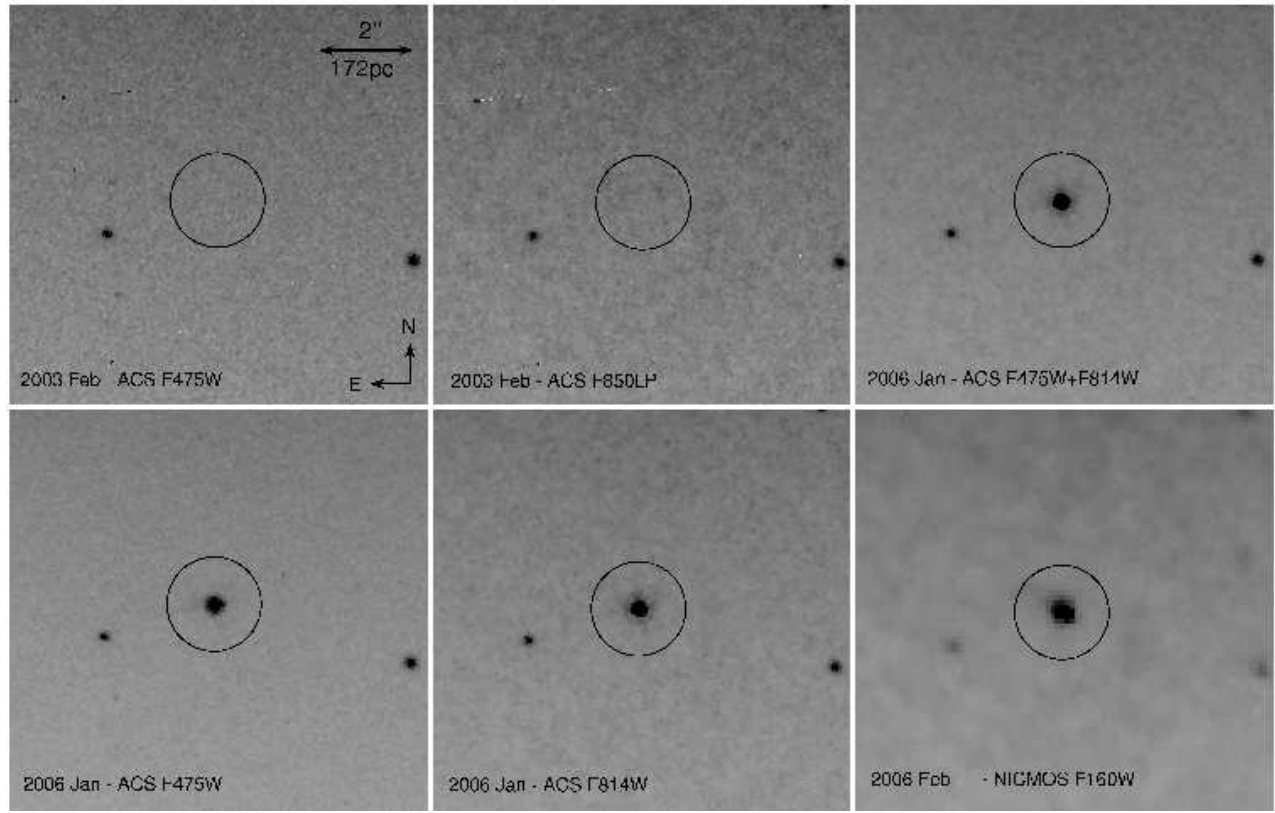}}
\caption{Zoom in on the environment of M85\,OT\,2006-1 HST/ACS and NICMOS/NIC2 images. The circle, with radius of $1''$, marks the position of the transient.
Note that the $F475W+F814W$ is a sum of the $F475W$ and the $F814W$ images.
\label{f2}}
\end{figure*}

%--- HST ACS ---
M85 was observed using HST/ACS on 2003 
as part of the HST-ACS Virgo Cluster Survey
(C\^ot\'e et al. 2004).
Subsequently, the transient was observed serendipitously
with ACS/WFC and NICMOS/NIC2 in 2006
(GO-10515) as a follow-up study to Peng et al. (2006).
The ACS observations on 2006 were obtained 18\,days
after the discovery of the transient.
The log of observations, the
measured magnitude of the M85 transient
or the $\sim3\sigma$ upper limit at the OT location,
as derived from the
HST images taken on 2003 and 2006,
are listed in Table~\ref{Tab-Mag}.
The HST images of the galaxy and the transient environment are
presented in Figures~\ref{f1} and~\ref{f2}.
\begin{deluxetable}{lclccc}
\tablecolumns{6}
%\tabletypesize{\scriptsize}
\tablewidth{0pt}
\tablecaption{Photometry and limiting magnitude}
\tablehead{
\colhead{Date} &
\colhead{Exposure} &
\colhead{Band} &
\colhead{Mag.\tablenotemark{a}} &
\multicolumn{2}{c}{Limiting mag.\tablenotemark{b}} \\
\colhead{} &
\colhead{s} &
\colhead{} &
\colhead{} &
\colhead{apparent} &
\colhead{absolute} \\
}
\startdata
2003 Feb 01   &  750 & $F475W (g) $  & \nodata & $>26.9$ & $>-4.5$  \\
2003 Feb 01   & 1120 & $F850LP (z)$ & \nodata & $>25.1$ & $>-6.2$  \\
%combined\tablenotemark{c}& 2954 & $F475W$ (g) & \nodata & $>27.1$ & $>-4.3$ \\
2006 Jan 24   & 2204 & $F475W (g) $  & $20.57$ &         &          \\
2006 Jan 24   & 2224 & $F814W (i) $  & $18.62$ & $>25.3$ & $>-6.0$  \\
2006 Feb 28   &  500 & $F160W$ (H) & $17.82$  & $>21.2$ & $>-10.1$ \\
\enddata
\tablenotetext{a}{Vega based magnitude corrected for infinite aperture (Sirianni et al. 2005). Errors in photometry are about $0.02$~mag for the ACS observations, and $0.05$~mag for the NICMOS observations. The NICMOS magnitude is calibrated using 2MASS stars in the field of view.}
\tablenotetext{b}{Vega based limiting magnitude as estimated by adding artificial point sources to the images in the neighborhood of the transient and inspection of the images for the added sources. The absolute magnitudes are calculated assuming a distance of $17.8$~Mpc to M85 (Mei et al. 2007)
and Galactic extinction of $E_{B-V}=0.031$ (Schlegel, Finkbeiner, \& Davis 1998; Cardelli, Clayton, \& Mathis 1989). Note however that distance estimates to M85 range between 14\,Mpc (Ciardullo et al. 2002) to 18.6\,Mpc (Blakeslee et al. 2001).}
%\tablenotetext{c}{Combined $F475W$-band images taken on 2003 and 2006.}
%\tablecomments{}
\label{Tab-Mag}
\end{deluxetable}
%

%--- HST NICMOS ---

%--- Spitzer ---

\begin{figure}
\centerline{\includegraphics[width=8.5cm]{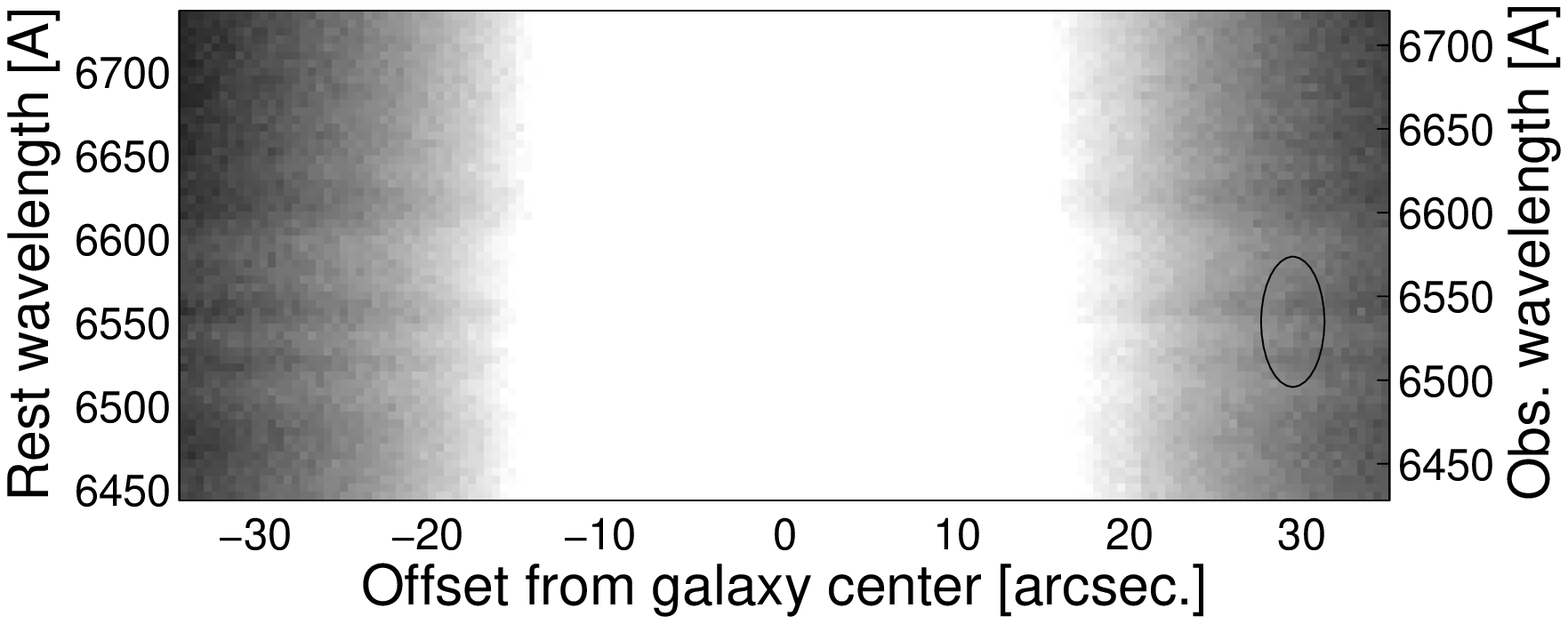}}
\caption{Two-dimensional spectrum of M85 and the transient
environment ($+30''$ offset from the galaxy center along the slit),
obtained about one year after the transient discovery, 
covering the H$\alpha$ wavelength region (marked by ellipse).
No H$\alpha$ emission
is seen in the vicinity of the transient. The spectrum is shown
before sky subtraction.
\label{M85_2d_spec}}
\end{figure}

%--- Palomar 200inch spectroscopy ---
On 2007 January 20, after the M85\,OT\,2006-1 faded away,
we obtained a spectrum
at the location of M85\,OT\,2006-1.
The spectrum (Figure~\ref{M85_2d_spec})
consist of $4\times300\,$s exposures
with the double beam spectrograph mounted
on the Palomar 5\,m telescope.
We used the $600$~lines/mm grating blazed at $9500\,$\AA~in
the red arm.
The $2''$ slit was centered on the nucleus of M85,
at a position angle of $185\,$deg.
The position angle was chosen such that
the location of the transient will be included
in the slit.

On 25 June 2005, M85 was observed by the Spitzer space telescope with the
Multiband Imaging Photometer for Spitzer (MIPS). 
The 70~micron image, with exposure time of 670~s,
is shown in Fig.~\ref{Spitzer_MIPS70}.

\section{Results}
\label{Res}

V838~Mon-like objects are found in both young regions
(e.g., V838\,Mon; Af{\c s}ar \& Bond 2007)
and old stellar populations (e.g., M31\,RV; Bond \& Siegel 2006).
M85\,OT\,2006-1 took place in an early-type galaxy.
Therefore, as we explain below,
it can be used to set an upper limit
on the minimal progenitor mass that can produce
V838~Mon-like eruptions.

\begin{figure}
\centerline{\includegraphics[width=8.5cm]{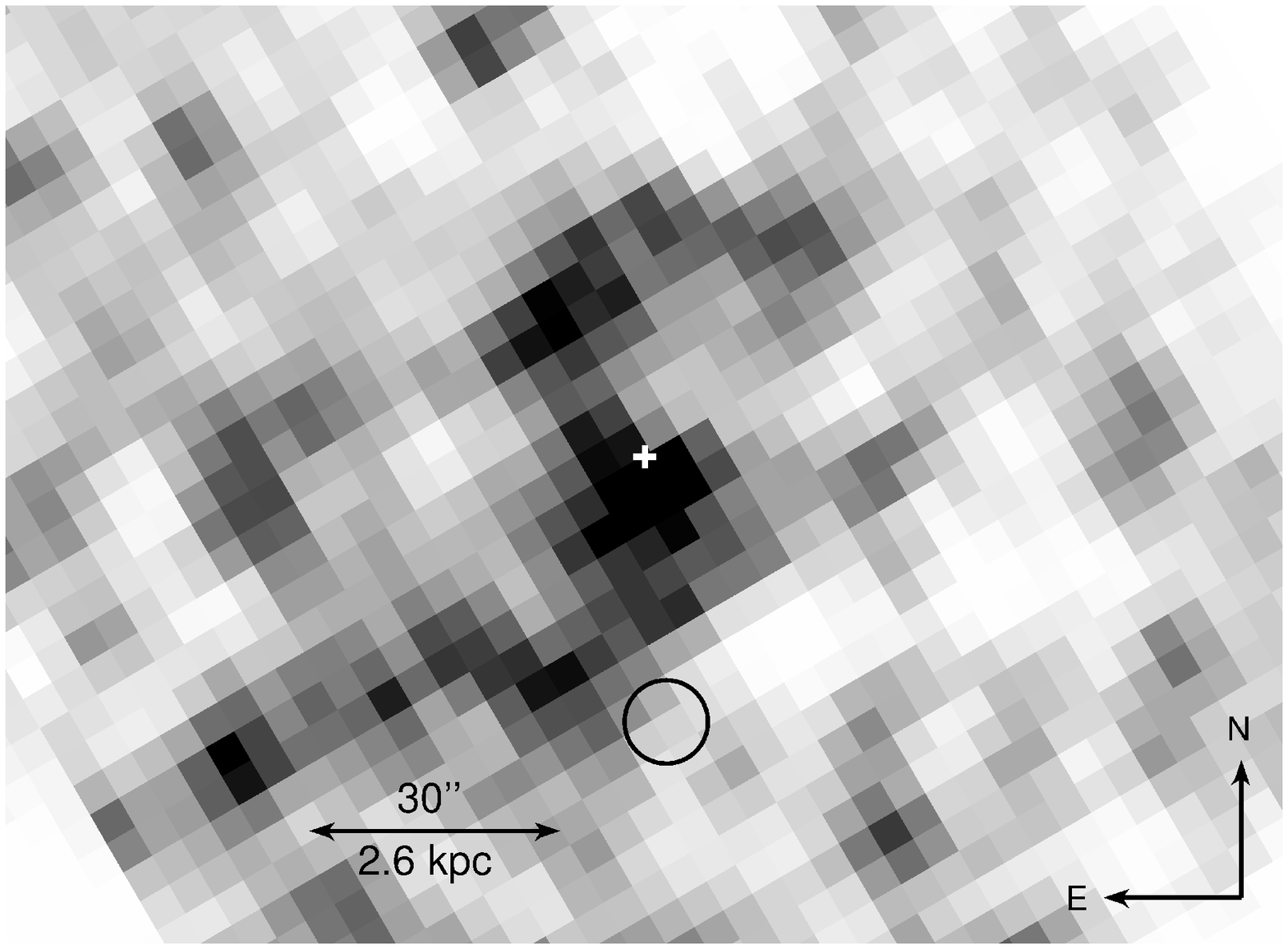}}
\caption{Spizter/MIPS 70 micron image of M85. The plus sign marks the
visible-light center of the galaxy, while the circle marks the position
of the transient.
\label{Spitzer_MIPS70}}
\end{figure}

From the 2-dimensional spectrum,
shown in Fig.~\ref{M85_2d_spec}, we can set an upper limit
on the flux of the H$\alpha$
emission at the location of M85\,OT\,2006-1 of
$<6\times10^{-14}\,$ergs\,cm$^{-2}\,$s$^{-1}$,
at the $3.5\sigma$ level
(equivalent to luminosity of $1.6\times10^{39}$~ergs~s$^{-1}$).
This corresponds to a star formation rate
smaller than about $10^{-2}$~M$_{\odot}$~yr$^{-1}$
(Kennicutt 1998) in a radius of $\sim100$~pc around the
transient location.
For comparison, the H$\alpha$ luminosity of the Orion nebula
is about $\sim10^{41}$~ergs~s$^{-1}$ (Haffner et al. 2003; assuming
a distance of 392~pc; Jeffries 2007).
Therefore, our observations rule out the presence
of a prominent star forming region in this location.
Moreover, based on the far-IR flux in the region of the transient,
obtained from the Spitzer/MIPS 70 micron image 
shown in Fig.~\ref{Spitzer_MIPS70}, we can set an upper
limit on the star formation rate in this region to
be less than $10^{-5}$~M$_{\odot}$~yr$^{-1}$ (Kennicutt 1998).
%
% Galaxy flux @ 70mu is ~25mJy -> ~1e-3 solar mass per year
% at position of OT, flux level is at least 100 times smaller...
%
The absence of \ion{H}{2} regions in M85 rule out
the possibility that the progenitor had a delay (from
birth to outburst) of $<10$~Myr (corresponds to $\gtorder40$~M$_{\odot}$),
which is the typical life time of \ion{H}{2} regions
(e.g., Mayya 1995).

An independent limit on the age and mass
of the progenitor can be inferred from
the absence of stars brighter than
$z$-band absolute magnitude $M_{z}=-6.2$
in the transient environment.
From the Lejeune \& Schaerer (2001) stellar tracks,
we find that stars older than 50~Myr
(and therefore, more massive than  7~M$_{\odot}$)
do~not reach $z$-band absolute magnitudes brighter than
$M_{z}=-6.2$~mag (at the red-supergiant stage).
We note that the $z$-band stellar track magnitudes were obtained
by interpolation of the $I$ and $J$-bands.
Therefore,
we can set a lower limit on the age of the most massive
stars in the transient environment to be $\gtorder50$~Myr, which corresponds
to mass $<7$~M$_{\odot}$ (assuming solar metallicity).
Otherwise,
we were likely to detect
individual stars in this region.
We note that we can limit the extinction in 
the transient location to $A_{i}<0.8$~mag, based
on the Balmer lines ratio, assuming case-B recombination
(Kulkarni et al. 2007).

%--- Color of the transient environment ---
%Furthermore, we compared the color-index of the environment
%of the transient as measured in the HST/ACS images
%to those of stellar spectral templates (Pickles 1998),
%and found that the color index of the environment
%is well matched with colors of K1 giants.
%We use this color to set a limit on the
%contribution of hot stars to the
%luminosity in the transient region, of less than about $5\%$,
%at the $3$-$\sigma$ confidence.

%--- Galaxy morphology ---
Ferrarese et al. (2006) reported
possible faint wisps and patches of dust in M85.
Moreover, Schweizer \& Seitzer (1992) reported that
M85 is somewhat bluer than typical
S0 galaxies, therefore possibly younger.
This claim is supported by Terlevich \& Forbes (2002)
who estimated the age and metallicity
based on line indices. They have found
a mean luminosity-weighted age of $1.6\pm0.3\,$Gyr and metallicity
of [Fe/H]$=0.44$ and [Mg/Fe]$=0.08$.
We note that the actual mean age is probably higher than that indicated by
line indices given that younger populations
have higher weight than old population.
%--- Age limits and progenitor mass limits ---
If the mean age is representative of the progenitor of
M85\,OT\,2006-1, then we can set a
lower limit of about $1\,$Gyr on
the age of M85\,OT\,2006-1 progenitor/s.
This further suggests that the mass
of the progenitor/s is probably below
$2$~M$_{\odot}$
(the life time of solar metallicity $>2$~M$_{\odot}$ stars is $<1$~Gyr;
Lejeune \& Schaerer 2001).
This limit is based on the mean stellar age
of this galaxy. However, stars younger than 1~Gyr
may be present in this galaxy in relatively small numbers.
%However, if such younger stars were the
%common progenitors of M85\,OT\,2006-1-like events then
%we would expect these transients to be discovered more frequently
%in late type galaxies, 
%rather than in environments dominated by old stars
%(e.g., M85; bulge of M31).

\section{Discussion}
\label{Disc}

Although several models exist for V838~Mon-like objects
(e.g., Soker \& Tylenda 2003; Lawlor 2005),
in the absence of detailed simulations, the nature
of these objects remain elusive.
A clue to their origin can be derived from
their environment, luminosity function and rate.
Given that only a small number of these
objects are known, and they were found serendipitously
in various searches,
the luminosity function and rate are not well constrained.
However, the fact that at least two events were observed in
our Galaxy 
(i.e., V838~Mon and V4332~Sgr)
in the last $\sim13$ years
suggests that they have a higher rate than SNe.
We can set a lower limit on
their rate, of $0.019\,$yr$^{-1}\,$L$_{MW}^{-1}$, at the $95\%$
confidence level, where L$_{MW}$ is the Milky Way luminosity.
%Given this rate,
%we note that if the mass ejected by these events
%(the amount of which is currently unknown) is considerable
%and rich in metals, then
%they may have some contribution to the chemical
%enrichment of galaxies.

%--- implications for models ---
Now we discuss the implications of our observations
for a specific model for V838~Mon-like objects.
Soker \& Tylenda (2006) presented a model for
violent stellar mergers
in which, prior to the merger, the spins and
orbital frequencies of the binary star are
losing synchronization due to the Darwin instability
(e.g., Eggleton \& Kiseleva-Eggleton 2001).
They found that for a given primary mass,
the maximal energy production obtained
for a binary mass ratio of $\sim1/50$,
is $\sim2.5\times10^{-3}GM_{1}^{2}/R_{1}$,
where $G$ is the gravitational constant,
and $M_{1}$ and $R_{1}$ are the mass and radius
of the primary star.
Given the upper limit on the progenitor mass, based on the mean stellar age
in M85, $<2$~M$_{\odot}$, and assuming a main-sequence mass-radius
relation, $R\propto M^{0.7}$, the maximum available energy
in their model is short by a factor of three in the total energy production,
as compared to the radiated energy
of M85\,OT\,2006-1 in the first two months, $\sim8\times10^{46}$~ergs
(assuming a distance of $17.8$~Mpc to M85; Mei et al. 2007).
Moreover, it is expected that a large fraction of the energy
will go into lifting the outer region of the star
rather than radiated away.
Furthermore, if the primary is an evolved star, then its radius
will be larger than the radius of a main sequence star
with the same mass, and the extracted energy will be even smaller.
This suggests that either more detailed modeling of
violent stellar mergers
is required, or that this event is not the result of a violent
stellar merger.
Another possible solution is that the mass of the progenitor is
somewhat larger. A larger progenitor mass will still be
consistent with our upper limit of $7$~M$_{\odot}$
which is based on the absence of stars brighter than $I\sim-6$~mag.
For example, according to Soker \& Tylenda (2006) model,
a $7$~M$_{\odot}$ progenitor can yield $\sim4$ times more energy than
a $2$~M$_{\odot}$ progenitor and may explain the discrepancy.
We note, however, that other kinds of instabilities
can lead to stellar mergers (e.g., in triple systems)
and that the above comparison is valid only
for the specific case discussed by
Soker \& Tylenda (2006).

Existing hydrodynamical simulations of the common envelope
phase in stellar mergers 
(and also star $+$ neutron star mergers)
predict that the total dissipated energy
is of the order of that observed
in V838\,Mon and M85\,OT\,2006-1 (e.g., Taam \& Bodenheimer 1989;
Terman et al. 1995; Terman \& Taam 1996).
Moreover, simulations of the common envelope phase predicts that
most of the envelope will be ejected in the
equatorial plane (e.g., Taam \& Ricker 2006).
Indeed, in Rau et al. (2007) we reported evidence suggesting that
the expansion of M85\,OT\,2006-1 is asymmetric.
However, more detailed hydrodynamical simulations of the
vast parameter space  available for stellar mergers are needed
in order to understand these processes and to test if V838\,Mon-like
objects are indeed the results of stellar mergers.

To summarize, we show that, in contrast to V838\,Mon,
but similarly to M31\,RV,
M85\,OT\,2006-1 was probably produced by members of an
old stellar population ($>1\,$Gyr),
and that its progenitor/s mass was probably
$\ltorder2\,$M$_{\odot}$.
%However, we can not rule out completly
%a progenitor with mass range of
%2 to 7~M$_{\odot}$.
These constraints narrow down the allowed 
venue of stellar models for the nature of this event.

\acknowledgments
%EOO thanks Orly Gnat
%for valuable discussions.
This work is supported in part by grants from NSF and NASA.

\end{document}